\documentclass[a4paper,twoside]{article}
\newcommand{\affil}[1]{$^{\rm #1}$}
\usepackage[authoryear]{natbib}
\bibpunct{(}{)}{;}{a}{}{,}
\usepackage{graphicx}
\date{} %Please leave the date blank
\newcommand{\etal}{\mbox{\it{et al}~~}}
\title{\large\bf\flushleft A Next Generation Deep 2$\mu$m Survey: Reconnoitering the Dark Ages}
\author{\parbox{\textwidth}{\flushleft
\vspace{-0.5cm}
{\it Jeremy Mould\affil{1,2}%, Second Author\affil{A,C}, and Third Author\affil{B}}\\
\vspace{0.4cm}
{\small \affil{1}Swinburne University,\ Hawthorn Vic 3122\\
 \affil{2}ARC Centre of Excellence for All-sky Astrophysics (CAASTRO)} 
\\
{\small Email: jmould@swin.edu.au}
}}}
\begin{document}

%SCAR AAA Meeting Sydney July 1, 2011
%\twocolumn[
\begin{changemargin}{.8cm}{.5cm}
\begin{minipage}{.9\textwidth}
\vspace{-1cm}
\maketitle
%
%
%%%%%%%%%%%%%     ABSTRACT    %%%%%%%%%%%%%
%Abstract of no more than 200 words here.
\small{\bf Abstract: The next generation 2$\mu$m sky survey should target nascent galaxies in the epoch of reionization
for spectroscopic followup on large telescopes. A 2.5 metre telescope at a site on the antarctic plateau has advantages
for this purpose and for southern hemisphere infrared surveys generally.}

%%%%%%%%%%%%%     KEYWORDS    %%%%%%%%%%%%%
\medskip{\bf Keywords:} infrared: general --- galaxies: high redshift --- telescopes
% Please write all keywords in lower case. PASA uses the
% standard list of subject headings adopted by The Astrophysical Journal
% and available from http://www.journals.uchicago.edu/ApJ/keywords_text.html.
% Keywords are separated by em-dashes, i.e. ---

%%%%%%%%DO NOT EDIT%%%%%%%%%%%%
\medskip
\medskip
\end{minipage}
\end{changemargin}
]
\small
%%%%%%%%EDIT FROM HERE%%%%%%%%%%%%

\section{Introduction}
%Please see the PASA Style Guide for help with correct layout for your manuscript.
%Examples of tables and figures are given below.

%The End of the Dark Ages: First Light and Reionization 
Until around 400 million years after the Big Bang, the Universe was a very dark place.  There were no stars, and there were no galaxies. Scientists would like to unravel the story of exactly what happened after the Big Bang. The PILOT survey telescope and the James Webb Space Telescope could pierce this veil of mystery and reveal the story of the formation of the first stars and galaxies in the Universe. 
Among other things the next generation infrared sky survey should target spectra and images of the first galaxies.
The JWST, GMT, TMT and European Extremely Large Telescopes need a source list. This can be provided by the PILOT Survey Telescope
(Lawrence \etal 2009).

An important current infrared sky survey is UKIDSS (Lawrence \etal 2007). That survey has covered
%The last decade state of the art in infrared sky surveys. 
7500 square degrees of the Northern sky, extending over both high and low Galactic latitudes, in JHK to K=18.3. 
It reached three magnitudes deeper than 2MASS (Skrutskie \etal 2006).
% UKIDSS = near-infrared SDSS
UKIDSS provides a panoramic atlas of the Galactic plane and is composed of five surveys 
There are two deep extra-Galactic elements, one covering 35 square degrees to K=21, and the other reaching K=23 over 0.77 square degrees.

The Current State of the Art is
VIKING - VISTA Kilo-Degree Infrared Galaxy Survey (Sutherland 2008).
The VIKING survey will image the same 1500 square degrees of the sky in Z, Y, J, H, and Ks to a limiting magnitude 1.4 mag deeper than the UKIDSS Large Area Survey. 
It will provide very accurate photometric redshifts, especially at z $>$ 1, an important step in weak lensing analysis and observation of Baryon Acoustic Oscillations. 
Other science drivers include the hunt for high redshift quasars, galaxy clusters, and the study of galaxy stellar masses.

The capacity of PILOT to go 2 to 3 mag beyond VISTA is considered in this paper, which concentrates
on high redshift galaxies. A full case would have a broader scientific focus. Testing an AST3 prototype (Zhao \etal 2010) for a survey class telescope has commenced.

\section{A PILOT Survey}% based on the PILOT telescope design}
A design study of an infrared 2 metre class telescope on the antarctic plateau was carried out by Saunders \etal (2008).%Telescope Design
Further details are provided by Lawrence \etal (2002a) and Storey\footnote{
http://www.aao.gov.au/pilot/pilot\_status.htm}, including the particular challenges of boundary layer
turbulence and water condensation. %PILOT was recommended for further study and for international
Other concepts have been discussed\footnote{ 
KDUST (http://www.kdust.org) and Ichikawa (2010).}.

Features of the PILOT design (Saunders \etal 2008) are: an Offner relay reflective cold stop design (diffraction limited),
on chip guiding, to beat read noise down by non destructive reads, 8K x 8K arrays giving a field 16$^\prime~\times$ 16$^\prime$  
at 0.125 arcsec/pixel.
The background at K is assumed to be 1mJy/sq arcsec, i.e. 14.54 mag.
The 0.2 arcsec aperture background would be K = 14.54 -- 2.5$\log(\pi 0.01)$ = 20.8 mag.
With a NICMOS sensitivity detector for H = 25, this gives S/N = 0.5 in 900s with background adjusted for aperture. 
To reach S/N = 2 requires 16 times longer, that is 4 hours.

The PLT is a European Polar Large Telescope and a 2$\mu$m survey would be an interesting basis for a European led collaboration. 
Their design incorporates a 40$^\prime$ field. %UNSW is a member of ARENA.
Equally, the proposed Japanese Dome F 2 metre infrared telescope or the planned Chinese Dome A 2.5 metre infrared telescope would be a fine choice for a survey.
The Australian peer review of PILOT advised that the project should be further studied but await international collaboration\footnote{http://astronomyaustralia.org.au/publications/ANSOC\_Report.pdf}. 
This now appears possible.

\section{Science Goals}
Although there are many science goals for a survey deeper than any previous one,
e.g. the lowest mass stars and star formation regions in our galaxy
(see also the ARENA (Burton \etal 2010) and Dome F proposals),
one of the most exciting is finding galaxies at redshift greater than 10 from the H dropout method.
This technique measures redshift by means of the Lyman break in the spectral energy distribution
of stellar populations.
These have no flux at 1.6$\mu$m, but are detected at 2.2$\mu$m.
The redshift of these is  1.6/0.09 -- 1 = 17.
Spectra of these objects would be obtained with JWST or a ground based extremely large telescope.

In this context the Antarctic advantages are
almost diffraction limited images over a wide field,
and low 2$\mu$m background. Lascaux \etal (2011) show median free atmosphere seeing of 0.23, 0.30 and 0.36
arcsec for Dome A, Dome C and the South Pole respectively.
This combination is only available from
the Antarctic plateau, high altitude balloons and space.
The best sites at temperate latitudes offer 0.8 arcsec seeing typically\footnote{http://www.gemini.edu/sciops/statistics\# weatherloss}.

The competition with PILOT is a space based survey: in particular, WFIRST (Levi \etal 2011).
WFIRST is focussed on exoplanets and dark energy.
The advantages of WFIRST are that it was top ranked in ASTRO 2010\footnote{
http://sites.nationalacademies.org/bpa/BPA\_049810}; it could deploy a
K broad filter; and there are
no clouds. The
disadvantages of WFIRST are that it has a
smaller aperture, 1.5 metre, and therefore a somewhat
lower resolution. There is only a
3 year mission lifetime and a proposed
2020 launch. Furthermore the cost would be
an order of magnitude higher than PILOT. The goal is
a 200 nJy limit $vs$ 70 nJy with PILOT.

Subaru (Ota \etal 2010) has found of order two Ly$\alpha$ emitters per square degree at z = 7.
With a similar $\lambda / \delta \lambda$ = 50 narrowband filter at K dark (Lawrence \etal 2002), PILOT could
conduct a similar search at z = 18. The constraint on the star formation rate at that epoch is direct,
but beyond the scope of the present paper on broadband surveys.

At z = 8 the Local Group would occupy some 3$^\prime$ (Wright 2006). If M31 is --20.3 mag in the blue
and it had a 20 times brighter stellar population at age 0.65 Gyr, it would show up at AB$_K~ \approx$ 26,
and K $\approx$ 24 mag, a solid detection in the proposed survey.

\section{The Dark Ages at 2$\mu$m}
At 2$\mu$m the dark ages extend from z = 6 when reionization is fairly complete 
to z = 2.4/0.09 -- 1 = 25, 
when the Lyman limit transits the long wavelength end of the window. 
WMAP has characterized 
 this interval as a time of strong, but anisotropic, star formation.
Figure 1 shows the duration of this period in time since the Big Bang and
the luminosity distance of sources at those redshifts. These immense distances are
a great challenge, but at least they protect observers from source confusion.
The lower panels show the K band evolution of stellar populations with instantaneous
and continuous star formation.   The single burst simple stellar population
is timed to go off at recombination. It ages and the Lyman jump passes
through the bandpass at high redshift, showing strong magnitude evolution for this reason. These models were constructed from Starburst99 data (Leitherer \etal 1999)with Salpeter IMF and solar composition and with stellar emission only.  
The continuous star formation model shows weak evolution, as we are seeing the emission from 
young stars which are continuously being replenished.

\section{Survey Bandpass}
As shown in Figure 2 the background varies rapidly across the 2$\mu$m window. The calculation
assumes 3.8\% mirror emissivity, $\epsilon_M$, which has been obtained on Gemini South, and zero atmospheric emissivity, $\epsilon_\lambda$.
Ashley \etal (1999) measured the background at an antarctic site. Understanding their results requires a detailed
model of $\epsilon_\lambda$ and the atmospheric transmission, T($\lambda$), which equals 1 - $\epsilon_\lambda$ according
to Kirchoff's Law. Water vapour opacity, dominating the short wavelength side of the window is the key component.
It is interesting to examine how signal to noise ratio varies with wavelength across the window for a flat spectrum
(per unit wavelength) source. SNR($\lambda$) $\propto$ w($\lambda$)/$\surd$ W($\lambda$), where w is a window function
and W is the Wien function for blackbody radiation. It is easy to show that d$\log$W/d$\log\lambda$ = h$\nu$/kT - 5.
$$w = (1 - \epsilon(\lambda))/\surd(\epsilon(\lambda) + \epsilon_M )$$
The value of $w$ is 5 when $\epsilon$ = 0, and the 10\% power points of $w$ (assumed here to be at 2.0 and 2.4 $\mu$m)
occur when $\epsilon(\lambda)$ = 0.5. At other wavelengths T = 1. This is the basis of the SNR calculation in presented in Figure 2.

Figure 2 suggests no clear basis for adopting a non standard K filter. All wavelengths contribute
to the overall signal to noise ratio more or less equally. 
The observations of Ashley \etal (1999) show a 2$\mu$m background that falls from 2.3 to 2.4 $\mu$m.
A more precise calculation would require atmospheric opacity modelling and inclusion of the OH emission background.
A partial mapping of that background is provided in Figure 3.
Matsumoto, Matsuura \& Noda (1994) find balloon based K sky brightness of 130 $\mu$Jy/arcsec$^2$.

\section{Survey logistics}
If one implemented a 20$^\prime$ field, the survey rate with PILOT is 26 years/sr,
assuming an unrealistic 180 $\times$ 24 clear hours per year.
However, Lawrence \etal (2002b) find a K background of 210 $\pm$ 80 $\mu$Jy/arcsec$^2$.
If this were adopted, the survey time would drop by a factor of 0.21$^2$ to a little over 1 year/sr.
Sky background is a major uncertainty. This necessitates further study both observationally
and by simulation.
The  PLT design offers 40$^\prime$ and therefore 6 years/sr.
A full survey would produce $\sim$100 Pb of data to do 2$\pi$ sr.
This is not a serious problem according to Moore's Law. Raw data
would be backloading for the observatory fuel supply.
Supernova monitoring would require computation and communication of a catalog in real time.

The focal plane for the PILOT telescope is ambitious and would be a good
basis for international collaboration. Optical alignment of 2 metre wide field telescopes
has proved challenging (e.g. Kaiser \etal 2010) and these lessons can be taken on board through collaboration.

PILOT would
not be obsolete until a proposed KDUST-8 metre telescope became operational.
That would reach AB$_K$ = 29 mag.
% What are we waiting for ?

\section{Conclusions}
Important targets at high redshift await the next generation 2$\mu$m survey.
These include galaxies with 10$^8$ year old stellar populations at redshift 6,
pair production supernovae from massive stars at M$_K$ = --23,
activity from the progenitors of supermassive black holes,
and young globular clusters with million year free fall times and mass to light ratios approaching 10$^{-4}$.
These are some of the inhabitants of the epoch of reionization that will surprise
us when we know where to point our spectrographs. There is time to test the AST3 prototype (Zhao \etal 2010) and build this telescope,
before narrow field adaptive optics corrected instruments such as 
GMTIFS\footnote{http://www.mso.anu.edu.au/gmtifs/infoday/docs/ gmtifs\_overview.pdf}
start to clamour for high redshift targets.

\section*{Acknowledgments} This paper is based on a presentation to the SCAR AAA meeting in July 2011.
I would like to thank the PILOT and PLATO\footnote{http://mcba11.phys.unsw.edu.au/$\sim$plato} teams for inspiration and for their perseverance and John Storey and his UNSW team for organizing a timely international meeting.
Survey astronomy is supported by the Australian Research Council through CAASTRO\footnote{www.caastro.org}.
The Centre for All-sky Astrophysics is an Australian Research Council Centre of Excellence, funded by grant CE11E0090.

%\end{multicols}
% It is preferable to embed your figures in the text as in the following example
%\pagebreak
%\begin{figure}[h]
%\begin{center}
%\includegraphics[scale=.7, angle=-90]{pilotfig1.eps}
%\end{center}
%\end{figure}

\pagebreak
\vskip 5 truein

\begin{figure}[h]
\begin{center}
\includegraphics[scale=.7, angle=-90]{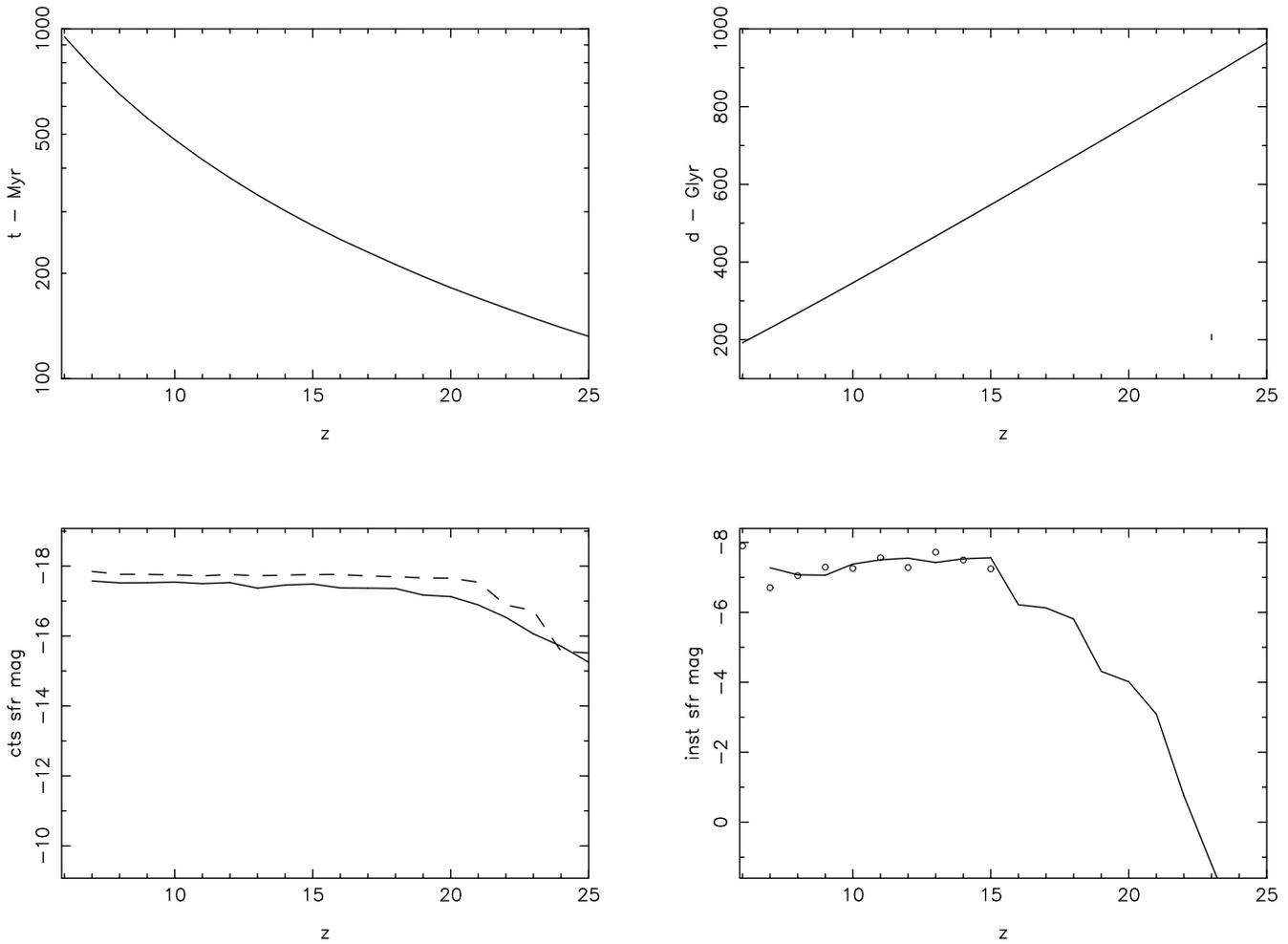}
\caption{The 2$\mu$m window is sensitive to epoch of reionization objects from as far as redshift 25.
Time since the Big Bang and luminosity distance in billions of light years are shown in the top two panels.
The remarkably small bar at z = 23, d = 200 Glyrs in the upper right panel is a Hubble radius.
The lower panels show the K band evolution of continuous star formation stellar populations (left)
and single burst stellar populations (right). The instant of star formation in the latter case was assumed
to be the epoch of recombination. The dashed line in the left panel is the result for metallicity one twentieth solar.
The open circles in the right panel are the result of direct look up of Leitherer's Starburst99 table. The solid
line results from interpolation in the table in age and integration over the K bandpass.}\label{figexample}
\end{center}
\end{figure}

\pagebreak

\begin{figure}[h]
\begin{center}
\includegraphics[scale=.7, angle=-90]{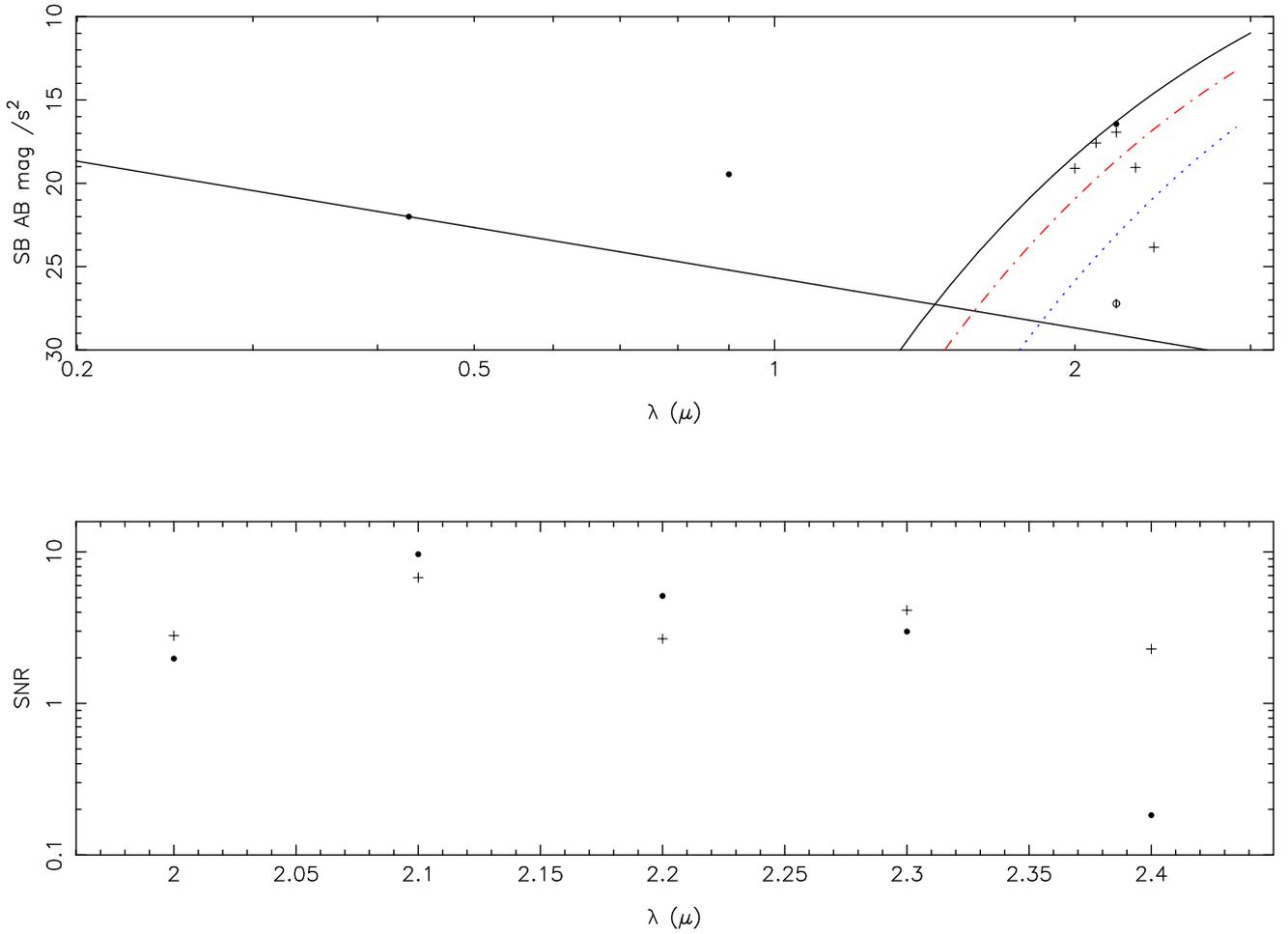}
\caption{The surface brightness of optical and near infrared photometric backgrounds is displayed in the upper plot.
In the optical the Rayleigh scattering wavelength dependence is normalized to B = 22 mag/arcsec$^2$.
In the infrared the solid curve is a 300K blackbody with emissivity 3.8\%. The solid symbol at 2.2$\mu$m is 1 mJy/arcsec$^2$.
The red (dot-dashed) curve is for 0$^\circ$C and the blue (dotted) curve for --40$^\circ$C. The measured K band cosmic background is shown as an open symbol with error bars (Wright 2001). In the lower plot the signal to noise ratio for a flat spectrum source observed
at --60$^\circ$C is plotted versus wavelength. Emissivity and transmission assumptions are described in the text.
An I band background of 19 mag/arcsec$^2$ is also shown; at this wavelength OH emission is dominant.
Plus signs show a recalculation using the OH intensity from Figure 3 normalized to a wavelength of 2$\mu$m.}
%The 2$\mu$m window }\label{figexample}
\end{center}
\end{figure}

\pagebreak
\begin{figure}[h]
\begin{center}
\includegraphics[scale=.7, angle=-90]{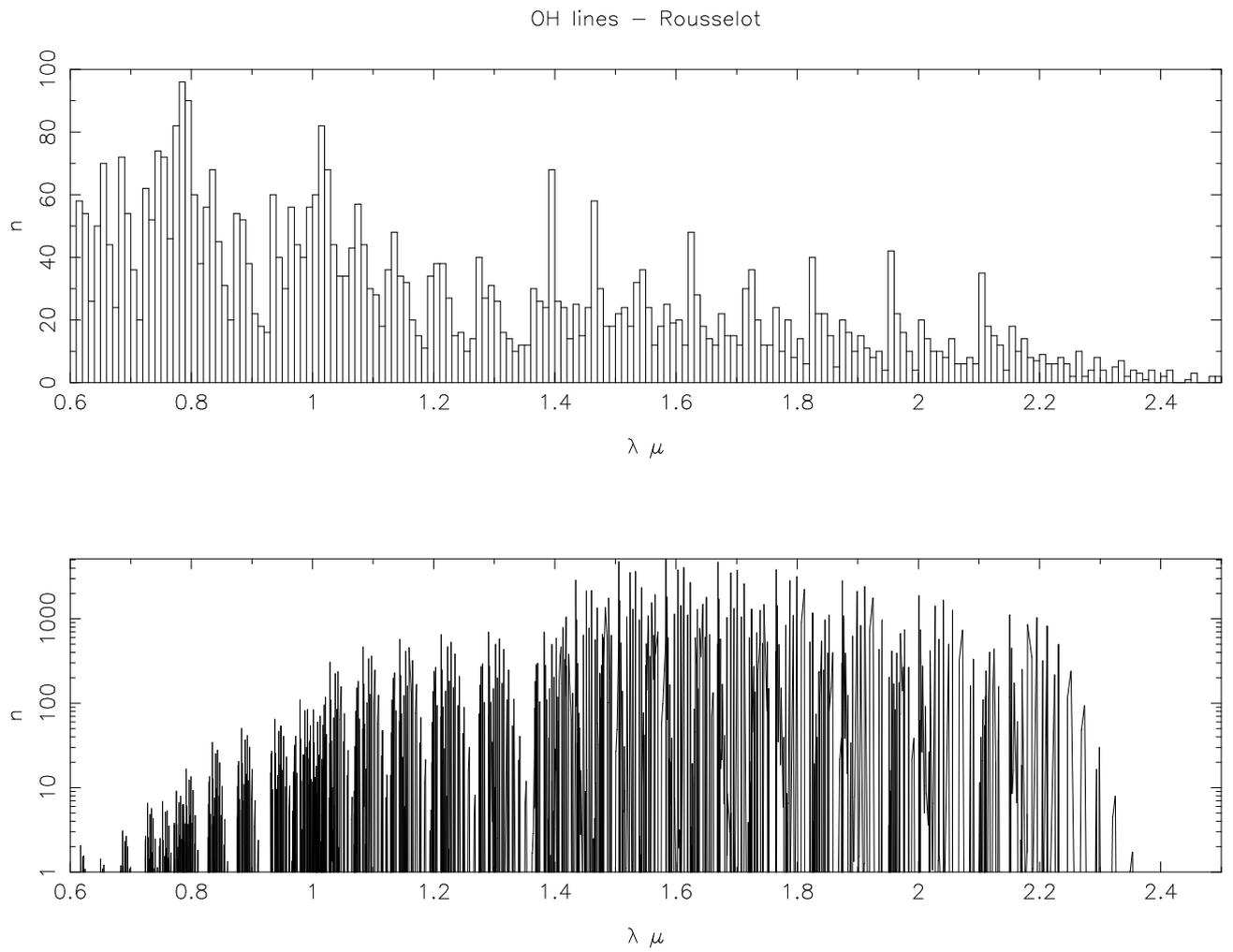}
\caption{$above$ The number density of OH lines from Rousselot \etal (2000). $Below$ the intensity.}
%The 2$\mu$m window }\label{figexample}
\end{center}
\end{figure}
%%Format tables as in the following example
%\begin{table}[h]
%\begin{center}
%\caption{Example Table}\label{tableexample}
%\begin{tabular}{lcc}
%\hline Column 1 & Column 2 & Column 3 \\
%\hline Table Content$^a$ \\
%\hline
%\end{tabular}
%\medskip\\
%$^a$Table footnotes go here.\\
%\end{center}
%\end{table}

\end{document}